# Legitimacy, Authority, and Democratic Duties of Explanation[1]

Seth Lazar, Australian National University

## I. Introduction

Here is my thesis (and the outline of this paper). Increasingly secret, complex and inscrutable computational systems are being used to intensify existing power relations and to create new ones; in particular, they are being used to govern (Section II). To be all-things-considered morally permissible new, or newly intense, power relations must meet standards of procedural legitimacy and proper authority. This is necessary for them to protect and realise democratic values of individual liberty, relational equality, and collective self-determination (Section III). For governing power in particular to be legitimate and have proper authority, it must meet a publicity requirement: reasonably competent members of the governed community must be able to determine that they are being governed legitimately and with proper authority. The publicity requirement can be satisfied only if the powerful can explain their decision-making to members of their political community. At least some duties of explanation are therefore *democratic* duties (Section IV). Section V applies these ideas to opaque computational systems, and clarifies precisely what kinds of explanations are necessary to fulfil these democratic values. Section VI addresses objections; Section VII concludes.

## II. Explanation, AI, Power

To explain X is to communicate information about X that enables some presumed audience to reach a justified understanding of X.[2] Our 'X' is acts—construed broadly to include decisions, verdicts, some omissions. Who should the bank deem creditworthy? Which social media posts should be removed? Who should receive a visa? What prompts will ChatGPT refuse to respond to? And so on. One can explain acts causally, describing for example the procedures that were followed or causal preconditions such as the option set available. Or one can give an explanation that invokes the agent's motivating or justifying reasons for acting, her beliefs or intentions when she acted.[3]

[1] This paper is forthcoming in *Oxford Studies in Political Philosophy*, and was presented to the Oxford Studies in Political Philosophy workshop, Tucson Arizona, October 2022. Many thanks to the organisers and other participants for their penetrating discussion. I presented versions at ANU, Carnegie Mellon, MIT and to the Michigan-Princeton AI Reading Group. Thanks to my hosts and to the audiences for their probing questions. For reading and commenting on drafts, particular thanks to: Christian Barry, Chris Bottomley, Garrett Cullity, Anne Gelling, Alex Guerrero, Alan Hájek, Atoosa Kasirzadeh, Todd Karhu (twice!), Niko Kolodny, Jeannie Paterson, Andrei Poama, Massimo Renzo, Pamela Robinson, Nick Schuster, Katie Steele, John Tasioulas, Peter Vallentyne, Bas van der Vossen, Kate Vredenburgh and Brian Weatherson.

[2] Daniel A. Wilkenfeld, 'Functional Explaining: A New Approach to the Philosophy of Explanation,' *Synthese* 191/14 (2014), 3367-91.

Our aim is a 'justified' understanding of X. The mere *feeling* of understanding is not enough. QAnon devotees no doubt *think* they understand American politics, but their understanding is not justified. Justified understanding is telic: it depends on the audience's goals. If I'm explaining how I built a Lego model of Hogwarts and the audience's goal is to build one themselves, then a justified understanding of my act requires an action-guiding causal explanation of how I did it. Justified understanding is also sensitive to the audience's capacities. An explanation of X that satisfies an expert may be impenetrable for a layperson.

So: to know what counts as an adequate explanation, we need to know why explanations matter and to whom they are owed. This paper aims to answer those questions for explanations of computational systems, especially those using Artificial Intelligence (AI). We rely on these systems in ever more spheres of our lives, but most of us do not know how they work, or why they yield the outcomes that they do. Their opacity (for our purposes, the antonym of explainability) derives from three sources.[4]

First, these computational systems are very often proprietary tools, kept *secret* from those affected by them. For example, the COMPAS algorithm used to inform pre-trial detention decisions in the US is the intellectual property of Northpointe and is secret.[5] The same is true for everything from DNA-matching algorithms widely deployed in criminal courts to the PageRank (Google) and Feed (Meta) algorithms that substantially govern our informational diets. Most notably, the new wave of 'foundation models' that are already radically reshaping the economy and may become the backbone of the next generation of platform capitalism all involve many layers of secrecy.[6] OpenAI will say neither

---

[3] Bertram F. Malle, *How the Mind Explains Behaviour: Folks Explanations, Meaning, and Social Interaction* (Cambridge, MA: MIT Press, 2004).
[4] The following draws from Andrew D. Selbst and Solon Barocas, 'The Intuitive Appeal of Explainable Machines,' *Fordham Law Review* 87 (2018), 1085-139; Jenna Burrell, 'How the Machine 'Thinks': Understanding Opacity in Machine Learning Algorithms,' *Big Data & Society* 3/1 (2016), 1-12.
[5] Julia Angwin et al., 'Machine Bias: There's Software Used across the Country to Predict Future Criminals. And It's Biased against Blacks' *ProPublica*, May 23 2016.
[6] Foundation models are very large AI models pre-trained using deep self-supervised learning so as to enable them to be subsequently fine-tuned for more specific purposes using supervised and reinforcement learning. Fine-tuned foundation models then become part of what I will call Generative AI Systems. 'Generative' AI is typically used to refer to AI systems that are able to generate content— text, code, images etc.–though in fact 'generative' is a technical term referring to a particular approach to classification in ML. See Rishi Bommasani et al., "On the Opportunities and Risks of Foundation Models," (2021), https://ui.adsabs.harvard.edu/abs/2021arXiv210807258B; Andrew Ng and Michael

what data GPT-4 (its most powerful Large Multi-Modal Model) was trained on, nor even how many parameters it has.[7] Many of the most important details that would help us understand GPT-4—not only how it works, but what its capabilities and limitations are—are simply kept secret. And one cannot reach a justified understanding of a secret.

Second, even when these tools are transparently deployed they are invariably too complex to be fully understood by any particular actor, because any given deployed product includes many different layers of technical and sociotechnical architecture. These features of computational systems are not new, but advances in AI over the last decade, in particular the rise of Machine Learning (ML), have introduced new sources of complexity that significantly exacerbate the explainability crisis.

As an example, consider the release of GPT-4; it was accompanied by a 100-page technical paper, and a 60-page 'system card' describing risks and mitigations.[8] Though these both shared important and valuable details about how the system was built and operationalised, they also make clear how complex the system is as a whole. Any given impact could be explained by the training data, the initial transformer-based self-supervised learning, the fine-tuning through supervised learning, the reinforcement learning with human and computational feedback, or the software architecture around the deployed version of GPT-4 in ChatGPT and the OpenAI API, with its additional layers of content moderation, security filtering, and other user design elements.

Or consider the computational architecture underpinning major social media platforms; they deploy complex systems involving layers of human content moderation, platform design, and amplification algorithms that integrate standard programming, ML models, user interface design and vast amounts of data, from both on-platform behaviour and from tracked behaviour online. The system as a whole has so many parts as to defy understanding by any individual inquirer.

Computational systems can in principle be arbitrarily complex, because no human mind needs to understand them in order for them to function effectively. Since there are often efficiency or compliance returns to adding additional elements to the technical architecture, and since their performance can be measured simply by monitoring inputs and outputs, increases in complexity are

---

Jordan, 'On Discriminative Vs. Generative Classifiers: A Comparison of Logistic Regression and Naive Bayes,' *Advances in neural information processing systems* 14 (2001).
[7] The parameters of a ML model are its internal elements that encode what it has learned from the training data. A larger number of parameters (other things equal) generally means a more powerful model.
[8] OpenAI, 'GPT-4 Technical Report,' (2023); OpenAI, 'GPT-4 System Card,' (2023).

inevitable.

Third, ML involves designing algorithms that learn from a body of training data, and create a model to enable predictions about data beyond the training set. Its success derives from the ability of incredibly powerful computational systems to derive patterns that are far more complex than human analysts could comprehend. ML models are often *inscrutable* to human analysts due in part to evincing mathematical properties that we find hard to understand.[9] They exhibit ultra high dimensionality: the models identify and weight the significance of relations among many different variables, representing potentially billions, in some foundation models trillions, of such relations.[10] In part due to this high dimensionality ML models often depart from smooth and comprehensible mathematical properties such as linearity, monotonicity, and continuity, exposing surprising jumps, changes in valence, and gaps. As a consequence, they often identify unintuitive and unexpected correlations. Most strikingly, foundation models trained simply to either predict a blacked-out token of text, or else to predict the next token, have demonstrated the capability to perform a wide range of downstream tasks that seem to have no obvious connection to the task they were trained on—including some level of mathematical reasoning, the ability to play chess, translate across languages, and most impressively to use other software tools to accomplish complex goals.[11] We simply do not know why these capabilities arise, and we cannot predict which further capabilities will arise with additional increases in scale.

The mathematical *processes* by which ML arrives at these inscrutable models are also inscrutable to both laypeople and the most advanced researchers. We can describe in general how a deep neural network operates, and what kinds of interventions are likely to lead to better performance against a set of benchmarks, but for any particular case we are reduced to radically empiricist methods: apply more GPUs and more data, and perhaps tweak the hyperparameters of the model until you get a result that performs better.[12] We don't know why it works—we just know that it does.

The opacity of these computational systems has sparked an extraordinary amount of research aiming to develop more explainable ML models, and to make fundamental scientific advances in understanding the internal operations of neural networks.[13] And many have proposed regulating this opacity—for example inscribing a right to explanation for automated decisions in European law.[14] But there is as yet relatively little substantive philosophical inquiry into precisely why explanations matter.[15] This is unfortunate. Both technical and regulatory work on explainability will succeed only if we have a clear sense of what counts as a good explanation! And to know what counts as a good explanation, we must know to whom explanations are owed, and why. To understand that, I argue that we must begin by recognising that these computational systems are not merely affecting our lives, they are creating new and intensified power relations. This does not unlock the whole truth about why explainability matters—we may sometimes care about explanations even when power is not being exercised. But it is an important start.

---

[9] Selbst and Barocas, 'The Intuitive Appeal of Explainable Machines'.
[10] Bommasani et al., "On the Opportunities and Risks of Foundation Models".
[11] Deep Ganguli et al., 'Predictability and Surprise in Large Generative Models' (paper presented at the 2022 ACM Conference on Fairness, Accountability, and Transparency, Seoul, Republic of Korea, 2022); Jason Wei et al., 'Emergent Abilities of Large Language Models,' *arXiv preprint arXiv:2206.07682* (2022); Timo Schick et al., 'Toolformer: Language Models Can Teach Themselves to Use Tools,' (Ithaca: Cornell University Library, arXiv.org, 2023).
[12] Gregory Wheeler, 'Machine Epistemology and Big Data,' in *The Routledge Companion to Philosophy of Social Science*, ed. Lee McIntyre and Alex Rosenberg (New York: Routledge, 2017), 321-29 This is true notwithstanding interesting advances in mechanistic interpretability, see e.g. Neel Nanda et al., 'Progress Measures for Grokking Via Mechanistic Interpretability,' *arXiv preprint arXiv:2301.05217* (2023).

Power is one-way control: the ability to shape others' prospects, options, and (evaluative and doxastic) attitudes without their being able to do the same to you.[16] Computational systems, especially AI, enable some to shape the prospects of others. Governments use AI to allocate healthcare and welfare, to track undocumented migrants, and to shape pre-trial detention decisions. Companies use AI to decide on individual creditworthiness, to price insurance, and to determine what products, services and content you are exposed to online. AI turns vast networks of CCTV cameras into inconceivably comprehensive and robust tools for mass surveillance. Algorithmic content moderation automatically flags potentially harmful speech at unthinkable scale across the internet.[17] And Generative AI Systems based on inherently inscrutable foundation models will automate many roles that previously involved accountable human decision-makers, from industry through to the public services.[18]

---

[13] Tim Miller, 'Explanation in Artificial Intelligence: Insights from the Social Sciences,' *Artificial Intelligence* 267 (2019), 1-38; Wes Gurnee et al., 'Finding Neurons in a Haystack: Case Studies with Sparse Probing,' *arXiv preprint arXiv:2305.01610* (2023); Nanda et al., 'Progress Measures for Grokking Via Mechanistic Interpretability'; Alex Foote et al., 'N2g: A Scalable Approach for Quantifying Interpretable Neuron Representations in Large Language Models,' *arXiv preprint arXiv:2304.12918* (2023).

[14] Sandra Wachter et al., 'Why a Right to Explanation of Automated Decision-Making Does Not Exist in the General Data Protection Regulation,' *International Data Privacy Law* 7/2 (2017), 76-99.

[15] The most notable exception is Kate Vredenburgh, 'The Right to Explanation,' *Journal of Political Philosophy* 30/2 (2022), 209-29. This gives an excellent justification for an individual right to explanation, but one orthogonal (and complementary) to my own argument, which focuses not on the rights of decision subjects, but on duties owed to the political community at large. For another argument grounded in rights of decision-subjects, see Reuben Binns, 'Algorithmic Accountability and Public Reason,' *Philosophy and Technology* 31/4 (2018), 543-56.

[16] See passim Keith Dowding, ed. *Encyclopedia of Power*, (Beverly Hills: SAGE, 2011) The following synthesises empirical research from a range of sources; for more detail, see Seth Lazar, 'Automatic Authorities: Power and AI,' in *Collaborative Intelligence: How Humans and AI Are Transforming Our World*, ed. Arathi Sethumadhavan and Mira Lane (Cambridge, MA: MIT Press, 2024), . And for a helpful overview see Jamie Susskind, *Future Politics: Living Together in a World Transformed by Tech*, First edition. ed. (Oxford: Oxford University Press, 2018). Some might reject my claim that power has to be one-way, arguing that A can have power over B, while B has power over A; this would change little in the ensuing argument.

Besides their direct impacts on our lives, computational systems also shape the options among which we choose. In our digital lives, this often means simply removing options dispreferred by the designer.[19] But they can also more subtly shape our choices: for example, 'persuasive technology', and 'dark patterns' whereby companies try to nudge us into choices that favour their interests (such as sharing more data than we might otherwise intend).[20] 'Opinionated' language models have already been shown to shape the beliefs of those using them.[21]

And of course, AI is the central organising principle of the information economy—the mediator that enables us to navigate the functionally infinite amount of information available at any given time. So it substantially shapes our evaluative and doxastic attitudes. From political debate to public health, from friendship and social mores to every aspect of the economy, our beliefs and desires are shaped by algorithms that use the most advanced techniques in AI—deep neural networks, large language models, reinforcement learning—to infer and shape what we want to see.[22]

Computational systems, especially those using AI, enable fewer people to achieve bigger impacts on a wider range of choices in the lives of more people. They increase the *degree,* the *scope*, and the *concentration* of power at stake. On the last point, consider again the COMPAS recidivism prediction algorithm.[23] In the past, no individual could influence bail decisions across multiple jurisdictions in the US except through the proper legislative and judicial processes. But the CEO of Northpointe can influence *many* such decisions; instructing their engineers to focus on one understanding of fairness rather than another (for example) could ramify across dozens of jurisdictions.[24] Software in general plays this centralising role: the more people come to rely on a given piece of software (AI or otherwise), the more it centralises power.

---

[17] Robert Gorwa et al., 'Algorithmic Content Moderation: Technical and Political Challenges in the Automation of Platform Governance,' *Big Data & Society* 7/1 (2020), 1-15; Tarleton Gillespie, 'Content Moderation, AI, and the Question of Scale,' *Big Data & Society* 7/2 (2020), 1-5.
[18] Tyna Eloundou et al., 'GPTs Are GPTs: An Early Look at the Labor Market Impact Potential of Large Language Models,' *arXiv preprint arXiv:2303.10130* (2023).
[19] Roger Brownsword, 'In the Year 2061: From Law to Technological Management,' *Law, Innovation and Technology* 7/1 (2015), 1-51; Susskind, *Future Politics*.
[20] Régis Chatellier et al., 'Shaping Choices in the Digital World, from Dark Patterns to Data Protection: The Influence of Ux/Ui Design on User Empowerment,' (CNIL, 2019)
[21] Maurice Jakesch et al., 'Co-Writing with Opinionated Language Models Affects Users' Views,' *Proceedings of the 2023 CHI Conference on Human Factors in Computing Systems* (2023), 1-15.
[22] For much greater detail on each of these dimensions of algorithmic power, see my Lazar, 'Automatic Authorities: Power and AI'; Seth Lazar, *Connected by Code: Algorithmic Intermediaries and Political Philosophy* (MS, 2023).
[23] Angwin et al., 'Machine Bias'.

And importantly, when power is sufficiently highly concentrated, even individually modest impacts can amount to a significant degree of power in the aggregate. For example, for any given individual the decision over whether to promote or demote their online communication might be relatively low-stakes. But a digital platform's decisions about algorithmic amplification at scale can have a decisive impact on the digital public sphere and its contribution to a functioning democratic society.[25]

Computational systems, especially those using AI, have intensified the power of those who already held it, and created new power relations, allowing some private companies to hold de facto dominion over great swathes of our lives. Crucially, these computational systems are often used to *govern* those subject to their influence. To govern is to settle on, implement, and enforce the norms that determine how an institution functions. When computational systems are used by government agencies in the exercise of their administrative or judicial functions and by social media companies to police the boundaries of online speech, or to determine our information diet, they govern us.

That these computational systems are secret, complex, and intrinsically inscrutable is clearly prima facie problematic. My task in the rest of the paper is to explain why.

III. Power, Legitimacy, Authority

Some critical studies of technology imply that merely to name power is enough to criticise it. But power need not be evil. It can protect the weak from the strong, scaffold inconstant individual agency, or indeed realise social justice. Let's grant that, as of now and on the whole, power exercised by means of opaque AI systems is not being used for justified aims. But suppose it were. Even then, we would still have cause for concern. The power of some over others may not be necessarily all-things-considered morally objectionable, but it is presumptively in tension with basic democratic values such as individual freedom, relational equality, and collective self-determination. Each value can be interpreted differently, and the ensuing discussion should be robust across most reasonable interpretations. However, for clarity, I will precisify them as follows.

I understand liberty as negative liberty or protection from wrongful interference

---

[24] Deborah Hellman and Kathleen Creel, 'The Algorithmic Leviathan: Arbitrariness, Fairness, and Opportunity in Algorithmic Decision Making Systems,' *Virginia Public Law and Legal Theory Research Paper* 2021/13 (2021).
[25] For further discussion, see Lazar, *Connected by Code*.

*and the risk of wrongful interference* by others.[26] Negative liberty contrasts with *positive* liberty, which prioritises the ability to make authentic choices between desirable options, and *republican* liberty, which prioritises not minimising the risk of interference but eliminating its possibility.[27]

The ideal of relational equality has deep roots, but it came to prominence in contrast with the philosophical focus on *distributive* equality.[28] Rather than focusing on the distribution of some good within a population, it describes an aspiration that we should live in a society where we recognise one another as moral equals, and where the institutions structuring our interactions reflect and support that equality.

Over time, societies collectively, and largely unintentionally, create and sustain social structures that affect our choices, making some things possible and others impossible, shaping our beliefs and desires. Collective self-determination is the process of reducing our subjection to heteronomous social structures that inadequately reflect our values. It involves jointly seizing the reins of our shared lives, so that we are not only formally equal, but we actually have positive political power to shape the shared terms of our social existence.

I call these *democratic* values because, in this world, democratic institutions are the only means by which all three will be realised. Moreover, democratic institutions *constitutively* enable relational equality and collective self-determination: they are not simply means to realise those values (as is I think the case for individual liberty). The institutions of democracy are many and complex, and I will not attempt a catalogue. And democracy is in practice always an imperfect ideal. But no other institutional arrangement has any realistic prospect of coming as close to (constitutively) enabling the fulfilment of these foundational values.

With these democratic values in mind, we can see why new power relations warrant suspicion and how they must be justified or resisted. Very simply put: if A has power over B, then B is subject to the risk of wrongful interference by A. So, power is presumptively in tension with negative liberty. If A has power *over* B, then they presumptively stand in hierarchical social relations, undermining relational equality. And if A has power over B, C, and D, then, presumptively, the society comprising [A, B, C, D] together are not collectively self-determining.

---

[26] Matthew H. Kramer, 'Liberty and Domination,' in *Republicanism and Political Theory*, ed. Cécile Laborde and John Maynor (Blackwell, 2008), 31-57. Note that Kramer's view of negative liberty is morally neutral, whereas I think we should not view justified interference as a limitation on people's negative liberty. Nothing substantial in what follows should change if you prefer to adopt a non-moralised conception of negative liberty.

[27] Philip Pettit, 'Freedom and Probability: A Comment on Goodin and Jackson,' *Philosophy and Public Affairs* 36/2 (2008), 206-20; Robert E. Goodin and Frank Jackson, 'Freedom from Fear,' *Philosophy and Public Affairs* 35/3 (2007), 249-65. My reasons for focusing on 'probabilistic negative liberty' as distinct from republican liberty concern both the impossibility and undesirability of eliminating the possibility of republican domination, and the tendency of the republican ideal of non-domination to conflate an internally diverse suite of normative concerns into a single value that has less explanatory power than separately focusing on (probabilistic negative) liberty, relational equality, and collective self-determination.

[28] Elizabeth S. Anderson, 'What Is the Point of Equality,' *Ethics* 109/2 (1999), 287-337; Samuel Scheffler, 'What Is Egalitarianism?,' *Philosophy & Public Affairs* 31/1 (2003), 5-39.

These presumptive objections can in theory be overridden if the exercise of power is *sufficiently* substantively justified. A good enough end *could* make illiberal, inegalitarian, authoritarian means all-things-considered permissible. But, first, these democratic objections would still leave a moral residue; second, this kind of justification applies only when one cannot achieve a comparably valuable end by less objectionable means. We therefore always have reason not just to override these objections, but to try to silence them. We do so by ensuring that power is used not only for the right ends, but *in the right way*, by those *with the right to do so*. I will call these three standards, respectively, substantive justification, procedural legitimacy, and proper authority, or the 'what', 'how', and 'who' questions.[29]

Contemporary analytical political philosophy often focuses too narrowly on state power, showing indifference to normative questions raised by power wherever it is found.[30] I think that *power over* is always presumptively objectionable, and so is always subject to these three distinct justificatory questions. However, sometimes the 'how' and 'who' questions are relatively easy to answer. For example, if A's power over B derives from the fact that B loves A, and A does not love B, then the legitimacy standard is maximally unconstraining, and A's authority derives from the simple fact that who she loves is entirely within her sphere of freedom.[31] Or else if the stakes are low, these standards can be very permissive. Undoubtedly procedural legitimacy and proper authority are most demanding for the power of the state over its subjects. But that is because of the stakes of state power, and the fact that the state governs its subjects. Whenever the stakes are comparably high, and power is used to govern, the legitimacy and authority standards should have some force.

What do legitimacy and authority look like when the stakes are high and power is used to govern? The fundamental idea of procedural legitimacy is to limit power by subjecting it to rules. This protects those subject to power against unwarranted interference in their prospects, options, and attitudes, as well as against the risk of such interference. Limiting power also restores some measure of relational equality, by giving us collectively the ability to rein in powerful individuals. *They* may have power over us in this decision, but *we* have power over them in ensuring that they act according to the standards that we have collectively set.

---

[29] Political philosophers often use the concepts of legitimacy and authority in confusing ways. My use of these terms is linguistically very simple. 'X exercises power legitimately means' that X exercises power in accordance with the constraints on that exercise of power. 'X has authority' means that X has the right to exercise power.

[30] Feminist political philosophy and the philosophy of race are obvious exceptions. See e.g. Iris Marion Young, *Responsibility for Justice*, ed. Samuel Freeman, Oxford Political Philosophy (New York: Oxford University Press, 2011); Charles W. Mills, *The Racial Contract* (Ithaca ; London: Cornell University Press, 1997).

[31] This is just one example. Other cases where procedural legitimacy seems to be less exacting include parental power, and power based on trust. But notice that in each case 'how' and 'who' standards clearly apply—they are just addressed in different ways than for the state.

I want to highlight three dimensions of procedural legitimacy. First, legitimately exercised power is limited in both range and degree, to the minimum needed to achieve its justified objectives. The powerful may exercise their power only in clearly defined ways, over a restricted sphere of activity.

Second, even when acting *intra vires* (within the bounds of their authorisation), the powerful must follow exacting procedural standards.[32] They must be guided by clear and comprehensible rules, which are publicly known in advance by those subject to them. Those rules should be applied consistently, without adverse or favourable distinction based on morally irrelevant features ('like cases should be treated alike').[33] There should be due process in the adjudication of claims, such that (for example) when one faces an adverse decision, one can see the evidence and reasons that support it, and mount a defence.

Third, power is exercised legitimately only if those in power are actually held to these standards through mechanisms of contestability and accountability, such that either the individuals adversely affected by their decisions, or we the people through our representatives, can challenge their decisions and ultimately replace those in power if they do not meet our expectations.[34]

Some believe the right to exercise power derives from nothing more than competence—any pro tanto objections to power's exercise are either silenced or overridden simply by using power wisely.[35] Others might argue that those who use power wisely and in the right way have a right to do so. I reject both of these views. Power must also be exercised by *the right people*: those with proper authority.[36]

There are many different ways to ground a right to exercise power, and they will vary depending on the nature of power in question. In what follows, I assume that the authority *to govern* should be grounded in *democratic authorisation*. Some have the right to govern others, because those others have authorised them to do so through democratic processes.[37]

---

[32] Cass R. Sunstein and Adrian Vermeule, 'The Morality of Administrative Law,' *Harvard Law Review* 131 (2018), 1924-78
[33] Jeremy Waldron, 'The Rule of Law and the Importance of Procedure,' *Nomos* 50 (2011), 3-31.
[34] Vredenburgh, 'The Right to Explanation'.
[35] Richard J. Arneson, 'Defending the Purely Instrumental Account of Democratic Legitimacy,' *Journal of Political Philosophy* 11/1 (2003), 122-32.
[36] Exercising power in the right way, for the right ends, may give one a strong *claim* to authority, but does not secure it in my view—in addition you also need to at least be *licensed* to exercise power by the broader political community.

Democratic authorisation matters because it constitutively enables relational equality and collective self-determination. A has power over the Bs. This undermines relational equality between them, and is presumptively antithetical to the Bs' collective self-determination. But if A's power over the Bs depends on the Bs' authorisation of A, then this restores relational equality, and by placing A in the effective control of the Bs, it enables them to be collectively self-determining.

IV. Legitimacy, Authority, Publicity

Computational systems are intensifying existing power relations and enabling new ones to be created. The stakes are often high, and these systems are being used to govern us. These novel power relations are presumptively morally objectionable and so must be justified against appropriate standards of legitimacy and authority. 'Explainability' matters because, in general, it is necessary for governing power exercised by means of computational systems to meet these standards.

My argument proceeds in two stages. In this section, I argue that for high-stakes governing power to be exercised legitimately and with proper authority it must satisfy the following *publicity* requirement: it should be possible for those who authorise that power's use to determine that it is being used legitimately and with proper authority. Then in the following section I show that explanations of computational systems are necessary to satisfy the publicity requirement.

A simple way to grasp the core idea of the first argument is just to imagine a state that exercised power in substantively justified ways but where it was strictly impossible for the citizens of that state to determine whether it was exercising power legitimately or with proper authority. It seems almost analytic that such a secret state could not meet those two standards.

Start with procedural legitimacy and recall its key components: ex ante limitation of what power can be used to do; in medias res constraints on precisely how power can be exercised; ex post contestability and accountability. At a minimum, the ex post constraints presuppose publicity: if we cannot tell whether the requirements of procedural legitimacy are being met, then we cannot hold the powerful accountable for not meeting them. But more than this, the ex ante and in medias res standards *should themselves* involve publicity requirements, since the values that procedural legitimacy is intended to serve are undermined in their absence. We authorise you to exercise power around here within bounds—one of which is that your exercise of power must meet a publicity standard. We limit your power by imposing constraints on precisely how you exercise it—one of those constraints is that you allow light into your decision-making processes. The value of procedural legitimacy is grounded (at least in part) in relational equality—the sense that while they have power over *us*, we have power over them by placing strict limits on how they exercise power. But relational equality is undermined if power is exercised in the dark.

---
[37] Daniel Viehoff, 'Democratic Equality and Political Authority,' *Philosophy & Public Affairs* 42/4 (2014), 337-75.

Social relations are social objects, constituted in part by how the people who inhabit those social relations understand them. If we cannot tell whether we are being treated as equals, then we do not enjoy egalitarian social relations. Of course, social relations also have an objective component—your mistaken belief that you are not being treated as an equal could not on its own undermine relational equality. But if you cannot tell whether you are being treated as an equal, then you are not.

If authority is grounded in authorisation, then it too entails a publicity requirement. Authorisation is structurally similar to consent (though it is more attenuated, institutionalised, and inherently collective). A's consent that B φs makes it permissible for B to do something that would be impermissible without A's consent. Likewise, when we authorise some to exercise power around here, we are making it permissible for them to do something that would be impermissible without our authorisation. Consent and authorisation are *morally effective* when they successfully enable this transformation of impermissible acts into permissible ones. This suggests three insights.

First, just as consent is dubiously morally effective when it is uninformed, the same is true for authorisation. Authorising someone to exercise power in secret is relevantly similar to consenting to someone's acting without knowing what you are consenting to.

Second, consent's moral effectiveness depends in part on its being public; the same is true for authorisation. This is clearly true for the moral effects of consent on third parties but is also plausible for the party whose otherwise-impermissible action is being consented to. If A 'consents' to B's sexual advance, without giving B any indication of their doing so, then it arguably remains impermissible for B to continue that advance. More generally, A's consent also changes the reasons that apply to others besides B. For example, suppose A consents to let B use A's car while A is out of the country. C, A's neighbour, sees B getting ready to drive off in A's car. Under normal conditions C would have reason to prevent what they perceive as a violation of A's property right. While A's consent to B taking the property objectively removes that reason, if A's consent is in no way public, then C still has reason, by his lights, to prevent B taking the property. So B should have some way of verifying his claim that A consented—A should communicate with C in advance, say, or give B some token (like the keys). Something similar is true for authorisation and authority. When the As authorise B to exercise power over them, that authorisation must be public, both to actually make it the case that B is permitted to exercise that power and to ensure that third parties know they have reason not to interfere.

Third, consent transforms impermissible acts into permissible ones; the withdrawal of consent can reverse that transformation. Publicity as a requirement on consent ensures that it remains current—that it has not been withdrawn. The same is true for authorisation. We can withdraw our authorisation for these people to exercise power over us. This relies on our authorisation being public so that we can know whether it is current, and effectively reverse it if necessary.

But authorisation is also somewhat different from consent. When we authorise some to exercise power over us, we not only make it permissible for them to do things they would not otherwise be permitted to do, we empower them to give us at least some content-independent reasons for action: we grant them authority over us. This too relies on authorisation being public. Suppose the residents of a town decide to deputise 100 new special litter constables. They use a computational system to select constables at random, and notify them directly of their being chosen. The system operates in secret, and no record is kept of the choice. The day after the selection, 100 new special constables are on our streets—but none of them can back up their assertions of authority to issue penalties for littering. Suppose you are confronted by one of these special constables, who enjoins you to pick up some litter nearby (which you did not in fact drop). Let's stipulate that if they had genuine authority, then the mere fact of their enjoining you to pick up the litter would give you some (defeasible) reason to do so and would also give third parties reason not to interfere in their exercise of authority. If you have no way of establishing their authority, have they given you or nearby third parties any kind of reason at all? Their authority over you is constituted in part by your knowledge (or reasonable belief) that they indeed have proper authority over you. This case therefore seems a failure of attempted authorisation. We have collectively tried to grant these special constables authority but because we deputised them in secret we have failed to do so.

Legitimacy and authority constitutively depend on publicity. To satisfy the publicity requirement it must be possible to determine whether power is being exercised legitimately and with proper authority. To do this, we need to understand how decisions were made and by whom. In other words, the powerful must be able to explain their decisions to we the people who authorise them to exercise power. The duty to explain decisions or decision systems is a duty of publicity. In the next section, I show how to apply this insight to computational systems. But first some preliminary observations.

If explainability duties are grounded in the publicity requirement, which itself is grounded in the values of legitimacy and authority, then explainability duties are owed to the same people who are owed legitimacy and authority. These values are, in turn, grounded in individual freedom, relational equality, and collective self-determination. Duties of explanation are therefore owed to the people whose individual freedom is constrained by those systems (which includes specific decision subjects), *but also (and primarily)* to the broader political community over whom governing power is being exercised, whose equality is at stake, and whose authorisation licences the exercise of power in this case.[38] In that sense, duties of explanation are *democratic* duties. They are owed to the democratic polity.

Consider a case in which power is exercised illegitimately or without proper authority, but with substantive justification. The individual subject to this decision *might* still be wronged by it—perhaps they have due process rights that have been infringed—but often substantive justification will be sufficient for the decision subject to lack any valid complaint against it. But the rest of us *clearly* have grounds for complaint against this illegitimate or unauthorised exercise of power. Illegitimate and unauthorised power wrongs all of us who collectively have a right to determine who exercises power around here and how.

This point merits emphasis. Explanations are necessary for publicity, publicity is necessary for authority. Proper authority matters especially because it is crucial for relational equality and collective self-determination. Democracy is the only institutional arrangement of governing power that has any prospect of jointly fulfilling those two values. Democracy is not a *means* to realise equality and collective self-determination. Successful democratic institutions themselves instantiate collectively self-governing relations of equality. So, explanations are necessary to realise the non-instrumental value of democracy.

If our duties of explanation are primarily understood as democratic duties, owed first to the political community, rather than (or as well as) to decision subjects, then the injunction is less to provide an explanation for every decision, more to ensure that those who exercise power can in general provide the political community with explanations, or resources from which to construct an explanation, for their decisions. Our goal is not the occurrent explanation of every decision but the possibility of providing such explanations if called on to do so.[39]

What's more, the publicity requirement might be equally well-satisfied by showing that *individual decisions* satisfy the legitimacy and authority constraints or by showing that *decision systems* do so.[40] The appropriate level of analysis likely depends on the stakes of the individual decision and the feasibility of providing explanations at a granular versus system level. For example, in cases where the stakes for individuals are relatively modest, but the aggregate stakes are high—as with the digital public sphere as described above—explanations are more urgent at the aggregate level.[41] In what follows I focus primarily on decision systems as being most relevant for establishing the legitimacy and authority of the exercise of power as a whole (rather than in particular cases).

---

[38] Scholars more commonly argue that explanations are owed primarily to the subjects of decisions. e.g. Margot E. Kaminski, 'The Right to Explanation, Explained,' *Berkeley Technology Law Journal* 34/1 (2019), 189-218; Vredenburgh, 'The Right to Explanation'.

[39] Thanks to Todd Karhu and Alex Guerrero for helping me to see this point.

Whether a given explanation enables a justified understanding of the explanandum inevitably depends on the epistemic capacities of the audience (as noted in Section II above). If duties of explanation are democratic duties owed primarily to the whole political community that authorises this exercise of power, then this shapes what counts as an adequate explanation. This does *not* imply that publicity, legitimacy, and authority depend on every one of us being spoon-fed an explanation for every decision that is tailored to our unique epistemic (in)capacities. Democratic citizenship places epistemic demands on us; these demands cannot plausibly or fairly be individually tailored irrespective of people's competence or effort. Instead, any reasonably competent member of the democratic community should be *able to* determine whether power is being exercised legitimately and with proper authority. The publicity requirement can be satisfied by explanations that enable a reasonably competent democratic citizen to determine that power has been exercised legitimately and with proper authority. This does *not* mean that we need to be able to understand absolutely every detail of the decision systems by which we are governed. The goal of explanation, on this argument, is strictly focused on the understanding necessary to know whether power is being exercised legitimately and with proper authority.

## V. Publicity, Explanation, AI

Publicity is partly constitutive of legitimacy and authority; for those who exercise high stakes governing power to satisfy the publicity requirement, they must be able to explain their decision systems to a reasonably competent citizen. When computational systems are used to exercise power, their opacity—due to secrecy, complexity, and inscrutability—makes it harder to explain the decisions to which they lead, and therefore undermines the publicity requirement and with it the legitimacy and authority of this exercise of power. But my aim here is not to issue a counsel of despair. Importantly, I do *not* think that we need detailed explanations of the technical operation of inscrutable and highly complex AI models in order to satisfy the publicity requirement. We *can* explain many important aspects of decision systems that use computational tools, including AI, and in doing so establish that the constituent elements of legitimacy and authority have indeed been satisfied. In this section I consider each of the constituent elements of legitimacy and authority and show how explaining different elements of decision systems that use computational tools like AI can help us determine whether those elements have been satisfied.

---

[40] Thanks to Finale Doshi-Velez here.
[41] See Lazar, *Connected by Code*.

Procedural legitimacy requires that significant decisions be made according to clear, defensible, publicly accessible rules. When computational systems support significant decisions, we must demand explanations of precisely which rules were being applied—and whether and how they were adapted to facilitate the computational approach.[42] Complex computational systems often bury the rules that they purport to apply or else apply rules that they have no business applying, simply because they can easily be implemented. For example, in the Australian 'Robodebt' scandal, an automated system sent out thousands of debt-collection notices to people it deemed had been overpaid benefits.[43] Its errors fell disproportionately on those who could least afford to suffer them. In the subsequent class action suit against the Australian federal government, it was revealed that the algorithm applied an 'income-averaging' rule that was explicitly deemed unconstitutional in the 1990s. An explanation of Robodebt's decisions showed that it applied rules it had no business applying. At present, GPT-4 makes governance decisions (such as whether to perform a particular action, or respond to a particular prompt) on entirely opaque grounds. Decisions might be the result of a content moderation layer, the underlying pre-trained model, or the fine-tuning in-between. They correspond only obscurely to OpenAI's vague terms of service. It would in principle be possible to provide a much clearer explanation for any given moderation decision than is currently afforded (Anthropic's publication of the 'constitution' for its competing model, Claude, is a positive step forward, though given the opacy of how Claude applies that constitution still leaves much to be desired).[44]

In criminal procedure, verdicts may be grounded only in admissible evidence—and not everything that bears on the truth of the verdict is admissible evidence. The same principle applies to procedural legitimacy more generally. We need to know whether decisions made by the powerful are based on appropriate evidence. For example, some kinds of data plausibly shouldn't influence certain kinds of decisions—your internet browsing history should not affect your creditworthiness, say, or the level of your insurance premium.[45] And some kinds of data should not be used to train ML algorithms—as in the case of ClearView.AI, which has built a facial recognition model on illicitly scraped data, which was never intended to be shared for that purpose.[46] Explanations of decisions made using computational systems should reveal the data on which the model was trained, allowing us to decide whether it really constitutes appropriate evidence for the decision at hand.[47]

---

[42] This doesn't matter only for legitimacy; it's also matters that people are able to adjust their behaviour to comply with the rules.
[43] https://www.smh.com.au/politics/federal/how-the-centrelink-debt-debacle-failure-rate-is-much-worse-than-we-all-thought-20170124-gtxh8q.html.
[44] Yuntao Bai et al., 'Constitutional AI: Harmlessness from AI Feedback,' (Ithaca: Cornell University Library, arXiv.org, 2022). See also https://www.anthropic.com/index/claudes-constitution.
[45] Barbara Kiviat, 'American Views on the Use of Personal Data in Two Market Settings,' *Sociological Science* 8 (2021), 26-47.

Algorithmic decision-making's propensity to mask or enable individual discrimination, and reproduce or exacerbate structural discrimination, is among its most widely remarked failings.[48] Procedural legitimacy demands that we treat (relevantly) like cases alike. To better understand whether this standard is being met, we can use counterfactual explanations for decisions, which hold morally relevant features of two decision subjects constant, while varying one that should be morally irrelevant, such as race.[49] Counterfactual explanations are not a panacea for structural discrimination.[50] But they can illuminate whether relevantly like cases have been treated alike, which is important for procedural legitimacy.

More generally, procedural legitimacy should protect us against *risk* of harm—by minimising both unjustified decisions, and *accidentally* justified decisions. When the correct decision is reached by proper procedures, we are not only treated fairly but are secure in that status. Indeed, explanations are strictly necessary for us *not* to be subjected to risk: if we do not know the process by which decisions that affect us are being made, then we must assign some substantial probability to their being made unreliably.

This guiding normative idea can help identify two further explainability goals. First, explanations must clarify whether the decisions were reached in a robust way—for example, would a minor perturbation in the input data have completely changed the outcome?[51] Were there multiple roughly equally well-performing models to choose from, which would have very different impacts on particular individuals, among which the engineer chose arbitrarily?[52] Would other optimisation rules, other measures of performance, or other tweaks to the model's hyperparameters have realised quite different results?[53] Probabilistic computational systems can often be alarmingly modally fragile, so these are realistic concerns. To protect us against the risk of bad decisions, we want the powerful to not just make the right decisions but to do so robustly—and explanations are necessary in order to assess the robustness of the decision, not just its accuracy.

---

[46] Ryan Mac et al., 'Surveillance Nation: A Buzzfeed News Investigation Has Found That Employees at Law Enforcement Agencies across the US Ran Thousands of Clearview AI Facial Recognition Searches – Often without the Knowledge of the Public or Even Their Own Departments.' *Buzzfeed News*, April 9, 2021 2021.

[47] For an attempt to establish norms of this kind, see Timnit Gebru et al., 'Datasheets for Datasets,' *arXiv preprint arXiv:1803.09010* (2018).

[48] Sam Corbett-Davies and Sharad Goel, 'The Measure and Mismeasure of Fairness,' *arXiv:1808.00023 [cs.CY]* (2018).

[49] Finale Doshi-Velez et al., 'Accountability of AI under the Law: The Role of Explanation,' *arXiv:1711.01134 [cs.AI]* (2017); Sandra Wachter et al., 'Counterfactual Explanations without Opening the Black Box: Automated Decisions and the GDPR,' *Harvard Journal of Law & Technology* 31/2 (2018), 841-87.

[50] In particular, ML systems are adept at inferring proxies for race, so a proper counterfactual test cannot involve either removing or reversing race classification in the model. This is in part because race is often so embedded in social identity that one cannot meaningfully hold other things constant while varying only that. See Issa Kohler-Hausmann, 'Eddie Murphy and the Dangers of Counterfactual Causal Thinking About Detecting Racial Discrimination,' *Northwestern University Law Review* 113/5 (2019), 1163-227. However, counterfactual explanations can sometimes be informative, as for example in the investigation into the COMPAS algorithm cited above.

Second, we want the powerful to make the right decision *for the right reasons*. For example, ML systems are excellent at inferring correlation but less adept at identifying causation.[54] Sometimes we need not only to predict whether you will suffer an adverse outcome, but whether that outcome will be your *fault* or not. If we cannot separate causation from correlation, then we cannot do this. Explanations can help us to see where correlations have been appealed to when causal claims were called for. More generally, we should, where possible, develop models for which we can identify the relative contribution made by different features (in isolation and combination) to a final verdict.

Procedural legitimacy also requires accountability. Complex computational systems make it easy to obfuscate human responsibility.[55] The risk is particularly great for tools using ML, since they are highly complex and are *supposed* to identify patterns that we cannot anticipate in advance. To serve accountability, explanations for decisions made using computational systems must surface the causal role of the people who actuated those systems. The other explanations referred to in this section have all aimed to identify the rules implemented by the computational system, the evidence on which it acts, the reasons (or features) that actuate it, the robustness of its responses. Accountability requires causal explanations: we need to clearly draw out the causal contributions of different human decision-makers to the outcome where the computational system decides this way or that.

Turn next to authority. Explanations are necessary for proper authority in at least two ways: explanations must *reveal* authorisation, and, when the authorised proxy acts on behalf of the principal, they must reveal why the proxy decides as it does. I expand on each point in turn.

---

[51] Anthony D. Joseph et al., *Adversarial Machine Learning* (Cambridge, United Kingdom ; New York, NY: Cambridge University Press, 2017).
[52] Amanda Coston et al., 'Characterizing Fairness over the Set of Good Models under Selective Labels,' in *Proceedings of the 38th International Conference on Machine Learning*, ed. Meila Marina (Proceedings of Machine Learning Research: PMLR, 2021).
[53] Umang Bhatt et al., 'Uncertainty as a Form of Transparency: Measuring, Communicating, and Using Uncertainty,' *AAAI/ACM Conference on Artificial Intelligence, Ethics, and Society (AIES)* (2021).
[54] https://www.technologyreview.com/2020/02/19/868178/what-ai-still-cant-do/
[55] Helen Nissenbaum, 'Accountability in a Computerized Society,' *Science and Engineering Ethics* 2/1 (1996), 25-42.

First, proper authorisation, like accountability, requires an audit trail. The specialist skills required to develop and deploy computational systems used to support government decision-making often lead to their being outsourced to subcontractors who clearly lack authority to adapt our laws in implementing them. In addition, our digital environment has grown faster than our capacity to regulate it and platforms often impose restrictions on their users without any democratic authorisation, pushing the boundaries of their authority over us. Explanations for computationally-supported exercises of governing power must therefore provide an audit trail which can show on demand that this decision was made by this agent, whose authority to make it was authorised by some other entity, all the way back to the sovereign authority of we the people.

For example, after nearly two decades of trying to figure out how to enforce intellectual property rights online, governments worldwide have outsourced enforcement of digital copyright to digital intermediaries, which are typically immune from liability for hosting pirated content provided they promptly prevent it from being viewed or shared on their platforms.[56] Accordingly, the major digital platforms have developed sophisticated algorithms to identify and remove content that may have been illegally shared. Their primary incentive is to minimise their costs and exposure to liability. So they predictably over-enforce. They are rewriting copyright law without any authorisation to do so: their exercise of power lacks authorisation. Explanations are necessary to reveal this.

Second, we need to understand the reasons for action of those whom we authorise to exercise power, at least when they act in our name, using our normative, political, and material resources to achieve common goals. When they do this, they represent us, and so their endorsement of some particular way of representing the world, or set of values, implies that we too endorse the same; we also are responsible for the things that they do, and the ways that they do them.

The design of computational systems with which to exercise power involves innumerable subtle value judgements. These evaluative decisions are buried when we focus only on the system's outputs as a whole; they must be surfaced through explanations for us to determine whether they should be rejected as a basis for public action on our behalf. Perhaps these reasons should be public in the Rawlsian sense of being, roughly, reasons whose validity as a basis for action on our behalf members of our community cannot reasonably reject.[57] But it probably matters more that they are a matter of public record, so that we can object to them if we want to.

---

[56] Nicolas P. Suzor, *Lawless: The Secret Rules That Govern Our Digital Lives* (Cambridge: Cambridge University Press, 2019); Gorwa et al., 'Algorithmic Content Moderation'.
[57] John Rawls, *Political Liberalism* (New York: Columbia University Press, 1993). For an application to the case of algorithmic accountability, see Binns, 'Algorithmic Accountability and Public Reason'.

In addition, recall that the value of authority is grounded, at least in part, in the value of self-determination. For a community to be self-determining, it should have *some* access to the reasons for action of those who exercise power on its behalf.

To see this, consider an analogy to individual self-determination. Imagine an individual who lives and dies by their horoscope, basing all their decisions on the gnomic pronouncements of their favourite astrologer. By chance, things actually go very well for them. Are they as self-determining as a counterpart who makes the same choices but actually has a justified understanding of those decisions and why they were the right ones to make? I think not. Understanding and endorsing why you are doing what you are doing, at least to some extent, seems to be an important contributor to individual self-determination. Some philosophers even think it is sufficient: that even if you cannot act otherwise, you are free so long as you act on reasons that you reflectively endorse.[58]

The same basic idea seems to apply to collective self-determination. If we have no idea why our proxy agents are making the decisions that they do, and so cannot reflectively endorse their reasons for doing so, we are to that extent heteronomous. Conversely, if we know why they act as they do, and we reflectively endorse their reasons for doing so, then that contributes to our degree of collective self-determination. Relying on computational models that even AI scientists cannot really understand is therefore in tension with genuine collective self-determination.

Of course, the world is a bleak and confusing place, and individuals and communities alike are often subject to forces that we don't control or understand. I do not claim that we are self-determining *only if* we can understand *everything* about our decisions and our lives. Only that intentionally relying on mystical or opaque processes to make our collective decisions leaves us *less* self-determining than we would otherwise be.

This argument further supports the call for explanations that (a) show that the computational system is being actuated by features that genuinely matter for the decision at hand—that it is 'acting for the right reasons'—and (b) demonstrate its robustness across various perturbations in the decision problem, and the training and test data.

---

[58] Harry G. Frankfurt, 'Alternate Possibilities and Moral Responsibility,' *The Journal of Philosophy* 66/23 (1969), 829-39.

VI. Objections

I explore two kinds of objections to my argument. The first series focus on whether explanations are really necessary to satisfy the criteria of legitimacy and authority. The second questions whether enough AI systems really exercise the kind of power that raises stringent demands of legitimacy and authority.

**Are Explanations Necessary for Legitimacy and Authority?**

1. An explanation of an act tells you how and why that act occurred. A *mere justification* explains (I hereby stipulate) the deontic status of an act, telling you why the act was (for example) permissible or impermissible. A *justifying explanation* explains the deontic status of an act, *as well as* explaining how and why the act occurred. Philosophers have long argued that mutual justification matters in political life, and that the exercise of power by the state should be justifiable to those affected by it. Is the publicity requirement really a public *justification* requirement? Can it be satisfied by providing mere justifications, rather than explanations?[59]

Mere justification cannot secure procedural legitimacy and proper authority; it answers only the substantive justification question. Even if power is used wisely, to do good things, it still constitutes a presumptive threat to our relational equality and collective self-determination if it is not used legitimately and with proper authority. We must care not only what decision was reached by the powerful, but how they reached it, and whether they had authority to make it. Mere justification for the decision itself does not address these questions.

Justifying explanations are more apposite, but have never been central to political philosophy. For example, Rawls explicitly cares only that coercive acts by the state should be *justifiable* by appeal to an overlapping consensus of reasonable comprehensive doctrines.[60] More importantly, why should only *justifying* explanations matter? The publicity requirement also applies when unjustified decisions are made; indeed we may need explanations then most of all.

2. When the stakes are high, many believe that explainability matters much less than accuracy. For example, if you had to choose between a medical treatment that could be properly explained and one that is proven to work better, though we do not know why, you would prefer the mysterious one that works better.[61]

This objection risks proving too much: one could make the same point about the exercise of power generally. Why care that power be used legitimately, as well as wisely? Because relational equality matters. Collective self-determination matters. Even if your dictator is wise and benign, you still have good reason to overthrow him just because he's a dictator. And in any realistic scenario, legitimacy serves accuracy—illegitimate power is unlikely to be used wisely, in the long run.

---

[59] See e.g. Binns, 'Algorithmic Accountability and Public Reason'.
[60] Rawls, *Political Liberalism*.
[61] Alex John London, 'Artificial Intelligence and Black-Box Medical Decisions: Accuracy Versus Explainability,' *Hastings Center Report* 49/1 (2019), 15-21.

Our intuitions about the case motivating this objection can be explained by its specific features, which are often absent from the scenarios being considered elsewhere in this paper. In the medical case, we can measure the accuracy of machine guidance over time using statistical studies and randomised controlled trials (RCTs). Additionally, in well-functioning healthcare systems, we have good reasons to trust our doctors, without monitoring their every decision, in part because the patient's and doctor's interests are aligned.

These three features of empirical verification, trust, and aligned interests are often absent from the exercise of power by means of computational systems. They often predict human behaviour in contexts too sensitive or complex for predictive models to be reliably verified with RCTs. Indeed, there may be no ground truth against which they can be measured, for example when a predictive algorithm indicates that a defendant is likely to be rearrested if released from pre-trial detention, and the judge therefore decides not to release him. In this case, there is no fact of the matter about what this defendant would have done if not released. RCTs could not conceivably be used in a situation such as this. And computational systems are deployed by public and private agencies in which we emphatically should not place our blind trust, and with which our interests are often not aligned. The counterexample therefore does not generalise widely—though it does offer some insight into when we can tolerate opacity.

3. Suppose we agree that explanations matter in principle. One might still think they are hard to come by in practice, not only for computational systems, but also for humans.[62] We never *really* know why humans reach their decisions. Our attempts at explanation are often post hoc rationalisations at best. If explanations are necessary for publicity, and so for legitimacy and proper authority, then so much the worse for us.[63]

This objection presupposes a depressing view of our capacity for rational decision-making, which I in general reject. What's more, the kind of explainability necessary for legitimacy and proper authority does not depend only on the luminosity of an individual's mental states. Explanations show how decisions were made: what procedures were followed; what evidence was used;

---

[62] E.g. Jon Kleinberg et al., 'Discrimination in the Age of Algorithms,' *Journal of Legal Analysis* 10 (2019).

[63] John Zerilli et al., 'Transparency in Algorithmic and Human Decision-Making: Is There a Double Standard?,' *Philosophy & Technology* 32/4 (2019), 661-83.

what rationale was presented; whether like cases were treated alike; whether decisions were made by those who were authorised to do so, and so on. Human decision-makers in institutional settings can explain their decisions by addressing these questions without analysing their private motivations. Algorithmic decision-making could in principle meet the same kinds of explanatory demands—the explainability crisis in AI has precipitated this debate by drawing attention to a moral phenomenon that was previously largely overlooked, but the kinds of explanations described in Section V are not beyond our technical capability, even now, provided we recognise that these kinds of procedural explanations matter and must not be obscured behind computational obfuscation or proprietary privilege.

4. Sometimes secrecy about the operation of computational systems seems necessary for them to function effectively. Consider algorithms at the heart of two-sided markets, search, and attention-allocation. If businesses knew how *those* algorithms work, they would be too easy to game.

This is a fair point, and sometimes the demand for explanations and its associated publicity requirement may indeed be overridden by other considerations. But the objection has greatest force against arguments for providing explanations to decision subjects—who will change their behaviour if they know how to game the system. I have argued, however, that explanations are owed to the political community, and this need can be served by providing explanations to our representatives, not directly to us.

As an example, consider the use of AI by a country's military against its adversaries. Obviously we wouldn't expect the military to explain AI-assisted decisions to those adversely affected by them, since the latter are our *adversaries*, and explanations would undermine our strategic objectives. However, the military should definitely explain those decisions to the representatives of the civilian population that it protects. It is acting on our behalf, in our name, with our stuff, and we have a right to know how and why it is doing so. These explanations should be provided in a secure environment, to democratic representatives whom we have entrusted with the oversight of these parts of our society. But the demand for explanations to ensure legitimacy and proper authority is by no means weaker for this.

### Does AI Really Raise Questions of Legitimacy and Authority?

I gave examples of AI systems being used in the exercise of power, and claimed that they are, moreover, used to govern. Some might question whether this is so, and argue that my exacting legitimacy and authority standards apply only infrequently to computational systems as they are actually deployed.

This is less an objection to my argument, than a challenge to its significance. My thesis can be restated as a conditional: if AI systems are used to exercise governing power, then they must meet the legitimacy and authority standards, and so those on whose behalf power is exercised are entitled to the informational resources necessary to determine whether those standards are met. I think that radical recent progress in AI research means that we will only see AI rule increase over the coming decade. In addition, I insist that computational systems *are* being used to govern us. Many of the highest stakes cases involve government use of AI. And we are increasingly governed by the technology companies that structure our digital lives—sometimes governing on behalf of states, sometimes filling a void left by state inaction.

Private companies are often the explicitly intended primary enforcers for statutory laws—most notably copyright law, or laws restricting harmful speech—that originate with more formal political institutions.[64] Whatever its flaws (and there are many), this is the central paradigm for governing the internet—even the regulations being advanced by the EU to regulate digital services and digital markets involve an extraordinary amount of delegation of governing power to private platforms. States outsource enforcement of the law to tech companies by creating significant penalties for companies that inadequately police their own platforms. Those companies then develop computational tools to enforce those laws—often leading to predictable over-enforcement in an effort to reduce liability.[65]

Moreover, many aspects of our digital lives are not adequately covered by statutory law (or else nobody is enforcing such laws at all, not even by proxy), and so private companies govern them de facto. The information age has generated new domains of social practice that desperately need to be governed—this is acutely clear in mid-2023, as deployments of Generative AI Systems are being explored in all sectors of economic and political life. Practically all of its most prominent ills—from disinformation to radicalisation, from surveillance and data extraction to febrile financial speculation—derive from coordination and collective action problems, or malicious actors, all of which can only be remedied by some centralised authority exercising governance power. These ills are as pressing and urgent as they are unlikely to be promptly addressed through statutory law. If private companies don't fill this vacuum with some kind of responsible approach to digital governance, then, at least for the near future, nobody else will. And given the scale of the task, as well as their obvious motivation to minimise the costs of governance, this invariably leads to their using secret, opaque, and intrinsically inscrutable computational tools to govern.

---

[64] Suzor, *Lawless*; Tarleton Gillespie, *Custodians of the Internet: Platforms, Content Moderation, and the Hidden Decisions That Shape Social Media* (New Haven: Yale University Press, 2018). Private companies have always been responsible for *applying* the law. But examples of their being the delegated authority to enforce the law by policing private citizens' behaviour are less common.

[65] Gorwa et al., 'Algorithmic Content Moderation'; Robyn Caplan, 'Content or Context Moderation? Artisanal, Community-Reliant, and Industrial Approaches,' (Data & Society, 2018); Gillespie, 'Content Moderation, AI, and the Question of Scale'; Suzor, *Lawless*.

Could one counter, here, that private companies' authority over us is grounded in our consent to their terms and conditions, and that they need meet only the procedural standards that they set out in those conditions? Of course, it is by now well-understood that our consent to digital services, like hypothetical consent, is not worth the paper it isn't written on.[66] But couldn't we engineer better models of more informed consent and so solve these problems of legitimacy and authority that way?

I can't of course rule this out, but I am sceptical.[67] One problem is simply that our consent to digital platforms has externalities for others (for example, through the data that we share, which enables inferences to be made about other people who do not consent to share their data). These externalities render consent dubiously morally effective, because we ourselves lack authority to sign others up to suffer the costs of our consent. Still more seriously, consent is morally effective only if you have a reasonable alternative to consenting. One can, of course, entirely opt out of the digital world—but this involves such significant personal costs as to not be a reasonable alternative. Consent no doubt has some role to play in developing legitimate and properly authorised structures of private digital power. But it cannot be the guiding or overarching principle, and it cannot negate the force of other considerations, such as the publicity requirement, and the concomitant duties of explanation.

VII. Conclusion

Public and private actors are using computational systems to govern us. With recent radical advances in AI capabilities, this will only increase. Unless necessary to achieve some extremely valuable goal, these new and intensified relations of governing power can be all-things-considered morally permissible only if they are procedurally legitimate and properly authorised. Legitimacy and authority constitutively depend on publicity: it must be possible for a democratic political community to determine that they are being governed legitimately and with proper authority. If it is not possible, then you already have your answer. Publicity requires explainability. In particular, the powerful must be able to provide members of the political community with explanations, tailored for the epistemic capacities of reasonable democratic citizens, that can establish whether decision systems satisfy procedural legitimacy and proper authority. I showed in Section V how the provision of specific kinds of explanations for computational decision systems can satisfy the publicity requirement with respect to each constituent element of legitimacy and authority. The ability to give technical explanations for how inscrutable AI models operate is not essential to the realisation of publicity.

---

[66] See e.g. Solon Barocas and Helen Nissenbaum, 'Big Data's End Run around Anonymity and Consent,' in *Privacy, Big Data, and the Public Good: Frameworks for Engagement*, ed. Julia Lane, et al. (New York: Cambridge University Press, 2014), 44-75.

[67] For further discussion of this line of argument see Lazar, *Connected by Code*.

Of course, if inscrutability, complexity, and secrecy are inherently in tension with procedural legitimacy and proper authority, then perhaps none shall 'scape whipping—benighted confusion and illegitimate power might be ineliminable features of the modern political condition. However, legitimacy and authority are not all-or-nothing properties. We are assessing highly complex systems; we cannot reasonably expect to reduce them to simple binaries. Legitimacy and authority admit of degrees, we can do better or worse with respect to each. Right now we are doing worse; we can do better.